\documentclass[11pt,draftcls,onecolumn]{IEEEtran}
\usepackage{graphicx}
\usepackage{caption}
\usepackage{subfigure}
\usepackage{amsmath}
\usepackage{amssymb}
\usepackage{bm}
\usepackage{color}
\usepackage{url}

\graphicspath{{./fig/}}

\begin{document}
\title{Application of Non-orthogonal Multiple Access   in LTE and 5G Networks}
 \markboth{\textit{A Manuscript Submitted to The IEEE     Communications Magazine} }{}
\author{Zhiguo Ding, Yuanwei Liu,  Jinho Choi, Qi Sun, Maged Elkashlan, Chih-Lin I \\and H. Vincent Poor
\thanks{
This paper was prepared in part under support of the UK EPSRC under grant number EP/L025272/1, and the U. S. National Science Foundation under Grants CNS-1456793 and ECCS-1343210.

Z. Ding and   H. V. Poor are with the Department of
Electrical Engineering, Princeton University, Princeton, NJ 08544,
USA. Z. Ding is also with the School of Computing and Communications, Lancaster University, LA1 4WA, UK.

Y. Liu and M. Elkashlan are with School of the Electronic Engineering and Computer Science, Queen Mary University of London, London, E1 4NS, UK.

J. Choi is with the  with School of Information and Communications, Gwangju
Institute of Science and Technology (GIST), Gwangju, 500-712, Korea.

Q. Sun and C.-L. I are with the Green Communication Research Center, China Mobile Research Institute, Beijing 100053, P.R. China.

  }
\vspace{-4em}}

 \maketitle

\begin{abstract}
As the latest member of the multiple access family, non-orthogonal multiple access (NOMA) has been recently proposed for 3GPP Long Term Evolution (LTE) and envisioned to be an essential    component of 5th generation (5G) mobile networks. The key feature of NOMA is to serve multiple users at the same time/frequency/code, but with different power levels, which yields a significant spectral efficiency gain over conventional orthogonal MA. The article provides a systematic treatment of this newly emerging technology, from its combination with multiple-input multiple-output (MIMO) technologies, to cooperative NOMA, as well as the interplay between NOMA and cognitive radio. This article also reviews the state of the art in the standardization activities concerning the implementation of NOMA in LTE and 5G networks.
\end{abstract}

\section{Introduction}
Non-orthogonal multiple access (NOMA) has been recently recognized as a promising multiple access (MA) technique to significantly improve the spectral efficiency of mobile communication networks [1-4]. For example, multiuser superposition transmission (MUST), a downlink version of NOMA,   has been proposed for 3rd generation partnership project  long-term evolution advanced (3GPP-LTE-A) networks \cite{MUST}. Furthermore, the use of NOMA has also been envisioned as a key component in  5th generation (5G) mobile systems  in \cite{Whitepaper1} and \cite{Whitepaper2}.

The key idea of NOMA is to use the power domain for   multiple access, whereas the previous generations of mobile networks have been relying on the time/frequency/code domain. Take the conventional orthogonal frequency-division multiple access (OFDMA) used by 3GPP-LTE as an example.   A main issue with this orthogonal multiple access (OMA) technique  is that its spectral efficiency is low when some bandwidth resources, such as subcarrier channels, are allocated to users with poor channel conditions. On the other hand, the use of NOMA enables   each user to have access to all the subcarrier channels, and hence the bandwidth resources allocated to the users with poor channel conditions can still be accessed by the users with strong channel conditions,  which significantly improves the spectral efficiency. Furthermore, compared to conventional opportunistic user scheduling which only serves the users with strong channel conditions, NOMA strikes a good balance between system throughput and user fairness. In other words, NOMA can serve users with different
channel conditions in a timely manner, which provides the possibility to meet the demanding 5G requirements of
ultra-low latency and ultra-high connectivity~\cite{Whitepaper1}.

In this article, we first provide an introduction to the basics of NOMA, such as typical NOMA power allocation policies, the use of successive interference cancellation (SIC) and the relationship between NOMA and conventional information theoretic concepts, where a simple example with two users is used to illustrate the benefit of NOMA. This introduction is then followed by a detailed overview of the recent developments in NOMA. We begin by considering the combination of NOMA and multiple-input multiple-output (MIMO) technologies. Various MIMO-NOMA designs will be introduced to achieve  different tradeoffs between reception reliability and data rates, since spatial degrees of freedom can be used to  improve either the receive signal-to-noise ratio (SNR) or the   system throughput.
The concept of cooperative NOMA will then be described, where employing user cooperation in NOMA is a natural choice since some users in NOMA systems know the information sent to the others and hence can be used as relays. In addition, cooperative NOMA has the potential to exploit the heterogeneous nature of future mobile networks, in which  some users might have better capabilities, e.g., more antennas, than the others. Therefore, the reception reliability of users with poor capabilities can be improved by requesting the ones with strong capabilities to act as relays. The interplay between NOMA and cognitive radio (CR) technologies, which have also been viewed as a key component of   next-generation mobile networks, will further be discussed, and  standardization activities to implement  NOMA in LTE and 5G networks will be also reviewed. Finally   research challenges and some promising future directions for designing spectrally and energy efficient NOMA systems will be provided, followed by concluding remarks.

\section{NOMA Basics}
\begin{figure}[t!]
    \begin{center}
        \includegraphics[width=0.6\textwidth]{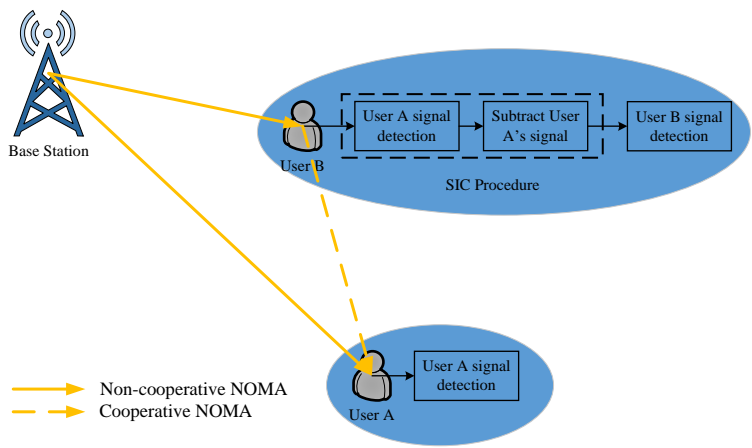}
        \caption{Illustration of a two-user NOMA network that involves  non-cooperative NOMA transmission without considering the cooperative phase illustrated by the dashed line.}
        \label{cooperative illustration NOMA}
    \end{center}
\end{figure}

In order to better illustrate the concept of NOMA, we take NOMA downlink transmission with two users as an example. As shown in Fig. \ref{cooperative illustration NOMA}, the two users can be served by the base station (BS) at the same time/code/frequency, but with different power levels. Specifically the BS will send a superimposed mixture containing two messages for the two users, respectively.  Recall that conventional power allocation strategies, such as water filling strategies, allocate more power to users with strong channel conditions. Unlike these conventional schemes, in NOMA,  users with poor channel conditions get more transmission power. In particular, the message to the user with the weaker channel condition is allocated more transmission power, which ensures that this user can detect its message directly by treating the other user's information as noise. On the other hand, the user with the stronger channel condition needs to first detect the message for its partner, then subtract this message from its observation and finally decode its own information.  This procedure is called SIC (as shown in Fig. 1).

The performance gain of NOMA over conventional OMA can be easily illustrated by carrying out high SNR analysis. With OMA, the achievable data rates for the two users are $\frac{1}{2}\log_2\left(1+\rho |h_A|^2\right)$ and $\frac{1}{2}\log_2\left(1+\rho |h_B|^2\right)$, respectively, where $\frac{1}{2}$ is due to the fact that the bandwidth resources are split between two users, $\rho$ denotes the transmit SNR, $h_A$ and $h_B$  denote the channel gains for User A and User B, respectively. Following Fig. \ref{cooperative illustration NOMA}, we assume $|h_A|^2<|h_B|^2$. At high SNR, i.e., $\rho\rightarrow \infty$, the sum rate of OMA can be approximated as $ \frac{1}{2}\log_2\left(\rho |h_A|^2\right)+\frac{1}{2}\log_2\left(\rho |h_B|^2\right)$. By using NOMA, the  achievable rates are $\log_2\left(1+\frac{\rho a_A|h_A|^2}{1+\rho a_B|h_A|^2}\right)$ and $\log_2\left(1+\rho a_B |h_B|^2\right)$, respectively, where  $a_A$ and $a_B$ are the power allocation coefficients. Therefore  the high SNR approximation for the NOMA sum rate is $\log_2\left(\rho |h_B|^2\right)$, which is much larger than that of OMA, particularly if the channel gain of User B is much larger than that of User A. In other words, the reason for the performance gain of NOMA is that the effect of the factor $\frac{1}{2}$ outside of the logarithm of the OMA rates, which is due to splitting bandwidth resources among the users, is more damaging than that of the factors inside of the logarithm of the NOMA rates, which are for power allocation.  It is worth pointing out that NOMA suffers some performance loss at low SNR, compared to OMA, if the strict NOMA power allocation policy is used.

 Downlink and uplink  NOMA can be viewed as special cases of multiple access channels (MACs) and broadcast channels (BCs), and therefore the rate regions achieved by  NOMA are bounded by the capacity regions   of the corresponding MACs and BCs. Compared to existing information theoretic  works which mainly focus on  the maximization of system throughput, a key feature of NOMA is to realize a balanced tradeoff between system throughput and user fairness.   Again take the two-user downlink case as an example. If the system throughput is the only objective, all the power will be allocated to the user with strong channel conditions, which   results in the largest throughput,  but User A is not served at all. The feature of NOMA is to yield a throughput larger than OMA, and also  ensures that users are served fairly. This feature is particularly important to 5G, since 5G is envisioned to support the functionality of Internet of Things (IoT) to connect trillions of devices. With OMA, connecting thousands of  IoT devices, such as vehicles in vehicular ad hoc networks for intelligent transportation,  requires thousands of bandwidth channels; however,  NOMA can serve these devices in a single channel use.   An important phenomenon in NOMA networks is that some users with poor channel conditions will experience low data rates. The reason for this is that these users cannot  remove their partners' messages completely from their observations, which means that they will experience strong co-channel interference and therefore their data rates will be quite small. In the context of IoT, this problem is not an issue, since many IoT devices need to be served with only small data rates.

\section{MIMO NOMA Transmission}

The basic idea of NOMA can be extended to the case in which a BS and users
are equipped with multiple antennas, which results
in MIMO NOMA. Of course, for downlink transmissions, the BS could use its multiple antennas either  for beamforming to improve the signal-to-interference-plus-noise ratio (SINR) \cite{Higuchi15}
or for spatial multiplexing
to increase the throughput \cite{Sun15}. We discuss these two options in the following sections.

\subsection{NOMA with Beamforming}

\begin{figure}[thb]
\begin{center}
\includegraphics[width=8cm]{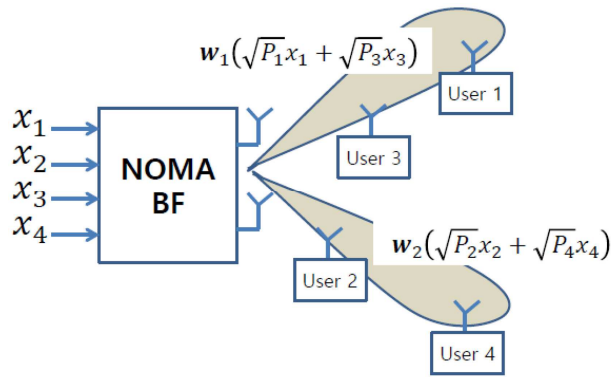}
\end{center}
\caption{An illustration of NOMA with beamforming.}
        \label{Fig:NOMA_BF}
\end{figure}

NOMA with beamforming (NOMA-BF) can exploit the power domain
as well as the spatial domain to  to increase the spectral efficiency by improving the SINR.
To see this, we consider a system of four users as shown in
Fig.~\ref{Fig:NOMA_BF}.
There are two clusters of users. User 1 and User 3 belong to cluster 1, while User 2 and User 4 belong to cluster 2. In each cluster,
the users' spatial channels should be highly correlated
so that one beam can be used to transmit signals to the users in the cluster.
For example, we can assume that ${\bf h}_3 = c {\bf h}_1$ for cluster 1,
where ${\bf h}_k$ is the channel vector from the antenna array at the BS
to user $k$, and for cluster 2, we have ${\bf h}_2 = c^\prime {\bf h}_4$, where $c$ and $c^\prime$ are  constants.
Furthermore, we assume that the beam to cluster 1 is orthogonal
to the channel vectors of the users in cluster 2, and vice versa.
That is, ${\bf w}_1 \perp {\bf h}_2, {\bf h}_4$ and
${\bf w}_2 \perp {\bf h}_1, {\bf h}_3$,
where ${\bf w}_m$ denotes the beam to cluster $m$.

Due to beamforming,
the signals from one cluster to the other   are suppressed. Thus,
at a user in cluster 1, the received signal would be a superposition
of $x_1$ and $x_3$, while
a user in cluster 2 receives a superposition
of $x_2$ and $x_4$, where $x_k$ is the signal to user $k$.
As shown in Fig.~\ref{Fig:NOMA_BF},
if User 3 is closer to the BS than User 1, User 3 would first decode
$x_1$  and subtract it to decode $x_3$ using SIC.
User 1 decodes $x_1$ with the interference, $x_3$.
Clearly,   conventional NOMA of two users
can be applied in each cluster.
In \cite{Zhiguo_mimoconoma}, this approach is studied to support
$2 N$ users in the same frequency and time slot
with $N$ beams that are obtained by
zero-forcing (ZF) beamforming to suppress
the inter-cluster interference.

A two-stage beamforming approach is proposed
using the notion of multicast beamforming in \cite{Choi15}.
In \cite{Higuchi15}, it is assumed that the users have
multiple receive antennas.
Thus, receive beamforming can be exploited
at the users to suppress the inter-cluster interference.
In this case, the BS
can employ a less restrictive beamforming approach
than ZF beamforming.

\subsection{NOMA with Spatial Multiplexing}

\begin{figure}[thb]
\begin{center}
\includegraphics[width=8cm]{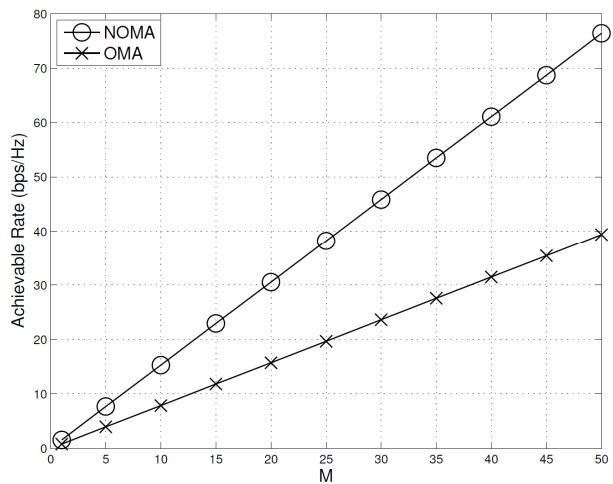}
\end{center}
\caption{Scaling property
of NOMA with spatial multiplexing.}
        \label{Fig:plt_CR4}
\end{figure}

Unlike NOMA-BF, the purpose of NOMA with spatial multiplexing (NOMA-SM)
is to increase the spatial multiplexing gain
using multiple antennas.  In NOMA-SM, each transmit antenna sends an independent data stream. Thus, the achievable rate can be increased by a factor of the number of transmit antennas.  This requires multiple antennas at the user as well.  This requires multiple antennas at the users.
In \cite{Sun15}, the achievable rate is studied
for NOMA-SM.
In principle, NOMA-SM can be seen
as a combination of MIMO and NOMA. Recall that the achievable rate of MIMO channels
grows linearly with the minimum of the numbers of transmit
and receive antennas under rich scattering environments, and therefore,
this scaling
property
of MIMO
should also be valid in
NOMA with spatial multiplexing.
Fig.~\ref{Fig:plt_CR4} shows
the achievable rate results of NOMA-SM
and OMA with  different number of antennas (denoted by M) under a rich
scattering environment.
It is assumed that the number of antennas
at the BS is the same as that at a user.
The power of the channel gain
of the weak user is four-times less than that of the strong user.
The total powers allocated to the strong and
weak users are 3 and 6 dB, respectively.
For OMA, we consider time division multiple access (TDMA) with equal
time slot allocation. Thus, each user's achievable rate in OMA
is the same as that of conventional MIMO.
However, since a given time slot is equally divided between two
users, each user's achievable rate is halved.
On the other hand, in NOMA,
each user can use a whole time slot and have a higher
achievable rate that could be two-times higher than that in OMA
as shown in Fig.~\ref{Fig:plt_CR4}.

\section{Cooperative NOMA Transmission}

\begin{figure}[!htp]\vspace{-1em}
\begin{center} \subfigure[Outage probability of the user pair]{\label{cooperative NOMA_system}
\includegraphics[width=0.47\textwidth]{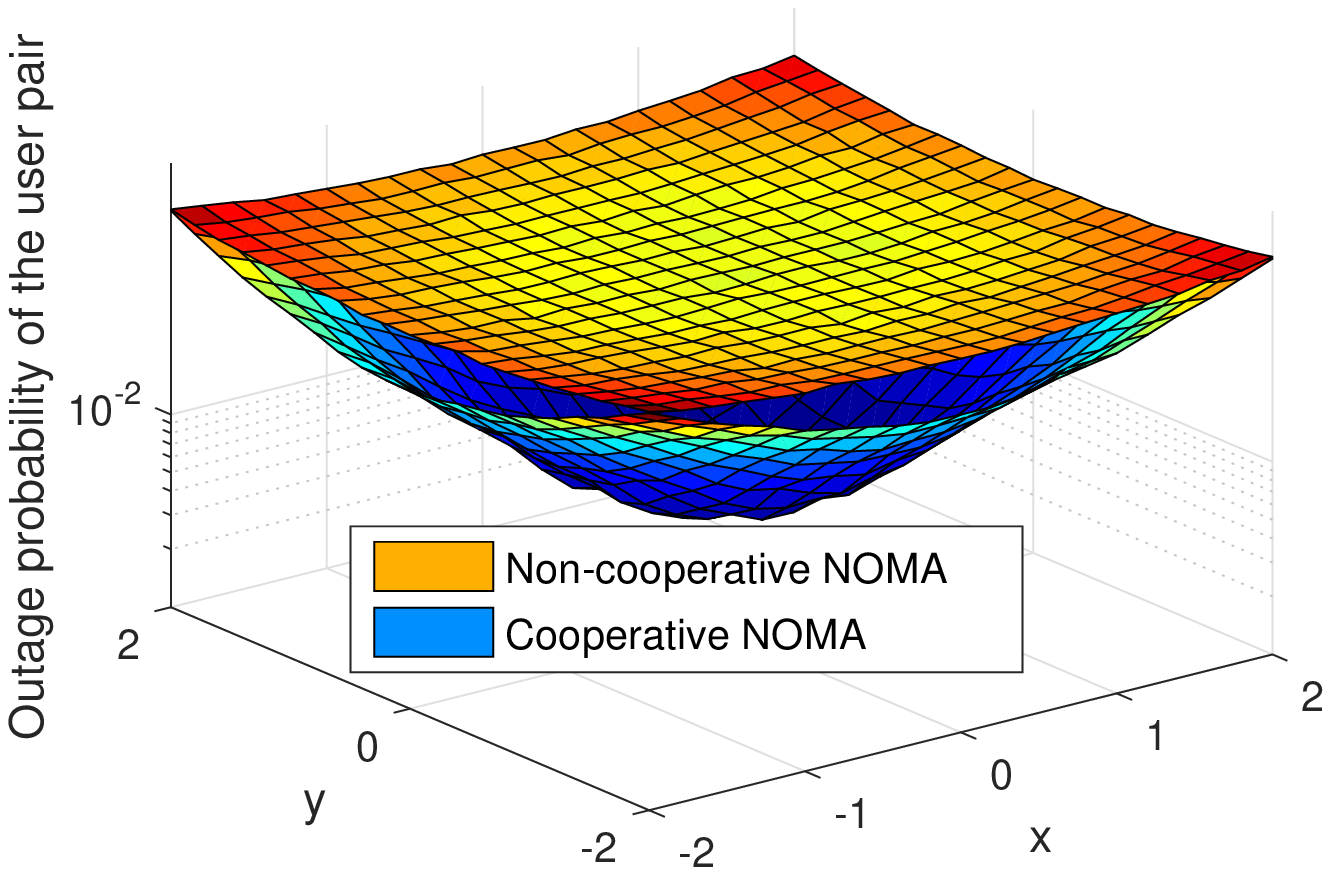}}
\subfigure[Outage probability of the poor user]{\label{cooperative NOMA_poor user}\includegraphics[width=0.47\textwidth]{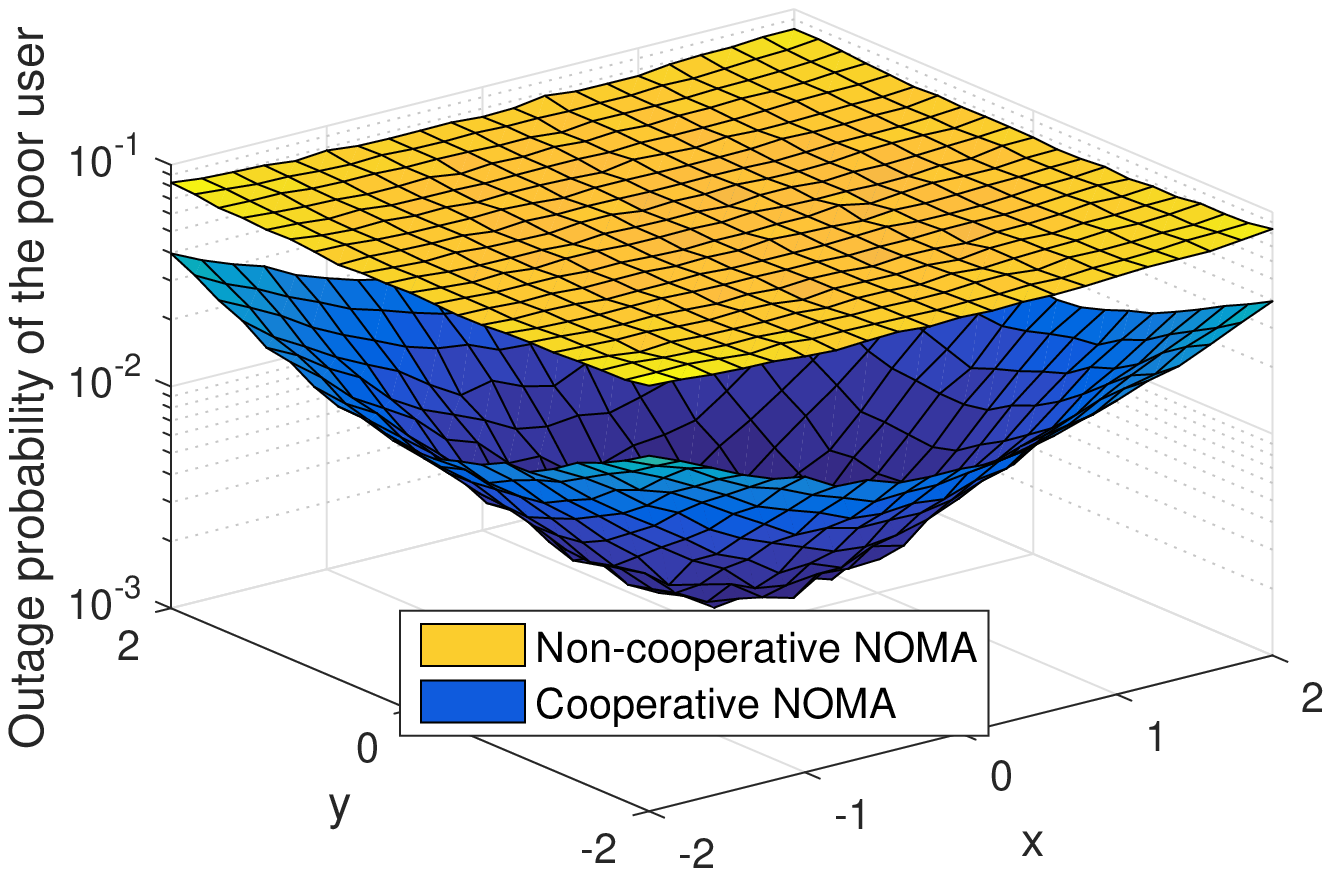}}\vspace{-1em}
\end{center}
  \caption{Performance of Cooperative NOMA transmission -- an example.  The BS is located at $(0,0)$. User A is located at $(5m,0)$. The $x$-$y$ plane denotes the location of User B. A bounded path loss model is used to ensure all distances are greater than one. The path loss exponent is $3$. The transmit signal-to-noise ratio (SNR) is $30$ dB. The power allocation coefficient for User A and User B are $(a_A,a_B)=\left(\frac{4}{5},\frac{1}{5}\right)$. The targeted data rate is $0.5$ bits per channel use (BPCU).}\vspace{-1em}
   \label{cooperative_NOMA}
\end{figure}

The basic idea of cooperative NOMA transmission is that users with stronger channel conditions act as relays to help users with weaker channel conditions. Again, take the two-user downlink case illustrated in  Fig.~\ref{cooperative illustration NOMA} as an example. A typical cooperative NOMA transmission scheme can be divided into two phases, namely, direct transmission phase and cooperative transmission phase, respectively. During the direct transmission phase, the BS broadcasts a combination of messages for User A (weaker channel condition) and User B (stronger channel condition). During the cooperative transmission phase, after carrying out SIC at User B for decoding User A's message, User B acts as a relay to forward the decoded information to User A. Therefore, two copies of the messages are received at User A through different channels. A more sophisticated and general  cooperative NOMA scheme involving $K$ users   was introduced in \cite{ding2014letter}. The advantages of cooperative NOMA transmission is that, since SIC is employed at receivers in NOMA systems, the messages to the users with weaker channel conditions have already been decoded by the users with stronger channel conditions. Hence it is natural to recruit the users with stronger channel conditions as relays. As a consequence, the reception reliability of the users with weaker channel conditions is significantly improved.  The performance improvement of cooperative NOMA  is illustrated in Fig.~\ref{cooperative_NOMA}. Particularly, Fig.~\ref{cooperative NOMA_system} and Fig.~\ref{cooperative NOMA_poor user} demonstrate that the cooperative NOMA outperforms non-cooperative NOMA in terms of the outage probability of the user pair and the outage probability of the poor user, respectively. In addition, in~\cite{ding2014letter}, it is demonstrated that cooperative NOMA  achieves a larger outage probability slope than non-cooperative NOMA, which is due to the fact that the former can achieve the maximum diversity gain for all users.


It is worth pointing out that complexity is an important consideration when implementing cooperative NOMA. For example,  it is not realistic to combine all users to perform cooperative NOMA. The main challenges are: 1) coordinating multi-user networks will consume a tremendous amount of system overhead, and 2) user cooperation will consume extra time slots. To overcome these issues, a hybrid MA system incorporating  user pairing/grouping has been proposed and viewed as a promising solution to reduce the system complexity of cooperative NOMA. Particularly, users in one cell can be first divided into multiple pairs/groups, then cooperative NOMA is implemented within each pair/group while OMA is implemented among pairs/groups. The performance of user pairing was investigated in \cite{Zhiguo_cr}, which demonstrates that pairing users with distinctive channel conditions yields a significant sum rate gain.

Furthermore,  it is important to point out that  power allocation coefficients have been recognized to  have a great impact on the performance of non-cooperative NOMA \cite{6692652}, and thus, investigating optimal power allocation to further improve the performance of cooperative NOMA systems is an important research topic.  There are other promising research directions based on cooperative NOMA. For example, considering simultaneous wireless information and power transfer (SWIPT), the NOMA user with the stronger channel condition can be used as an energy harvesting relay to help the user with the weaker channel condition without draining the latter's battery. A class of cooperative SWIPT NOMA protocols is proposed in \cite{yuanwei2014} and its performance is evaluated by applying  stochastic geometry.


\section{Interplay between Cognitive Radio and  NOMA}
NOMA can be viewed as a special case of CR networks. For example, consider a two-user scenario shown in Fig. \ref{cooperative illustration NOMA}.  User A can be viewed as a primary user in a CR network. If OMA is used, the orthogonal bandwidth allocated to User A cannot be accessed by other users, despite the fact that User A has a poor connection to the BS, i.e., the bandwidth resource allocated to this user cannot be used efficiently. The use of NOMA is equivalent to the application of the CR concept. Specifically User B, a user with stronger channel condition, is introduced to the channel occupied by User A. Although User B causes extra interference at User A and hence reduces User A's rate, the overall system throughput will be increased significantly since User B has a strong connection to the BS, as can be observed from Fig. \ref{fig set comparison b1}.

\begin{figure}[!htp]\vspace{-1em}
\begin{center} \subfigure[Fixed power allocation]{\label{fig set comparison
b1}\includegraphics[width=0.47\textwidth]{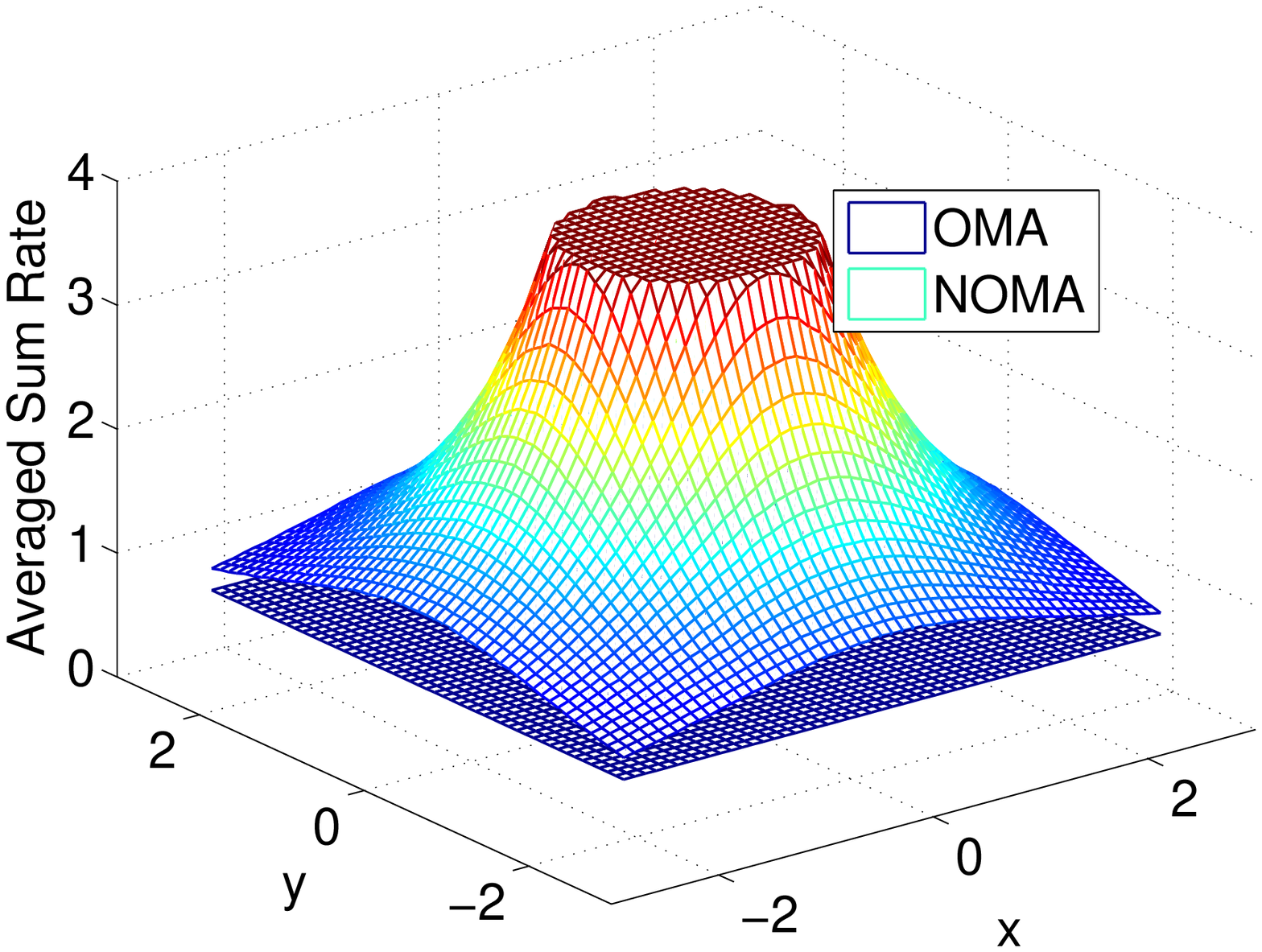}}
\subfigure[Cognitive radio power allocation]{\label{fig set comparison
b2}\includegraphics[width=0.47\textwidth]{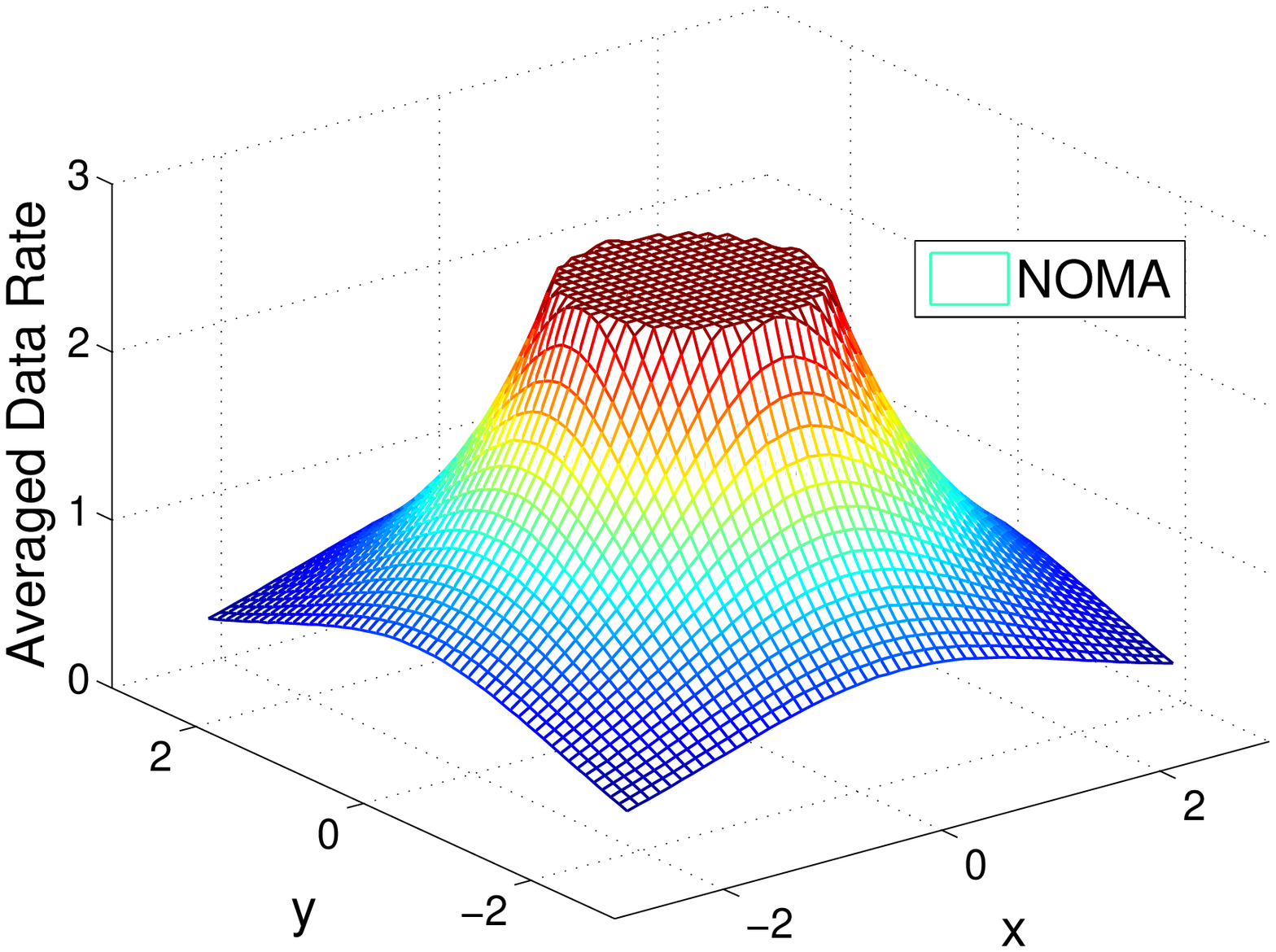}}\vspace{-1em}
\end{center}
  \caption{Performance of downlink NOMA transmission -- an example.    The transmit SNR is $20$ dB. A fixed power allocation policy, $(a_A,a_B)=\left(\frac{7}{8},\frac{1}{8}\right)$, is used in the first subfigure. The power allocation coefficient used in the second figure needs to satisfy a targeted data rate of $0.5$ BPCU at the primary user. The other parameters are set the same as in Fig. \ref{cooperative_NOMA}.   }\vspace{-1em} \label{figure2}
\end{figure}

The analogy with cognitive radio   not only yields insight into the performance gain of NOMA, but also provides guidance for the design of a practical NOMA system~\cite{Zhiguo_cr}. For example, NOMA seeks to strike a balanced tradeoff between system throughput and user fairness. However, user fairness can be measured by many different metrics. By using the CR concept, an explicit power allocation policy can be obtained to meet the users' predefined quality of service (QoS).  An example of such CR inspired NOMA networks is illustrated in the following by considering the two-user case shown  Fig.~\ref{cooperative illustration NOMA}. Consider that User A, i.e., the user with weaker channel condition, has a targeted data rate $R_1$. Here,  the CR inspired power allocation coefficient $a_A$ needs to satisfy $\log_2\left(1+\frac{a_A^2 |h_A|^2\rho}{(1-a_A^2|h_A|^2 ) \rho+1}\right)\geq R_1$.  The aim of this CR inspired power allocation policy is to ensure that the QoS requirements at the primary user, i.e., User A, are strictly met, and the BS can explore the degrees of freedom in the power domain to serve User B opportunistically, as shown in Fig.~\ref{fig set comparison b2}. It is worth pointing out that this CR inspired NOMA is particularly useful in the MIMO scenario, where it is difficult to order users according to their channel conditions and hence challenging to find an appropriate power allocation policy~\cite{Zhiguo_mimoconoma}.


The interplay  between cognitive  radio and NOMA is bidirectional, where NOMA can also be applied in CR networks to significantly increase the chance of secondary users to be connected. For example, without using NOMA, separated bandwidth resources are required to serve different secondary users, which can potentially introduce a long delay for secondary users to be served. The use of NOMA can ensure that multiple secondary users are served simultaneously, which   effectively increases the connectivity of the secondary users. Power allocation at the secondary transmitters is critical to the application of NOMA in CR networks. Specifically, it is important to ensure that the secondary users are served without causing too much performance degradation at the primary receiver, where the total interference observed at the primary receiver is an important criterion.  Furthermore, the power control policy used also needs to ensure that interference among the secondary users is carefully controlled in order to meet the secondary users' QoS requirements.

\section{State of the Art for NOMA in 3GPP LTE and 5G}
\begin{figure}
  \centering
  \subfigure[MUST Category 1]{
  \includegraphics[width=0.5\textwidth]{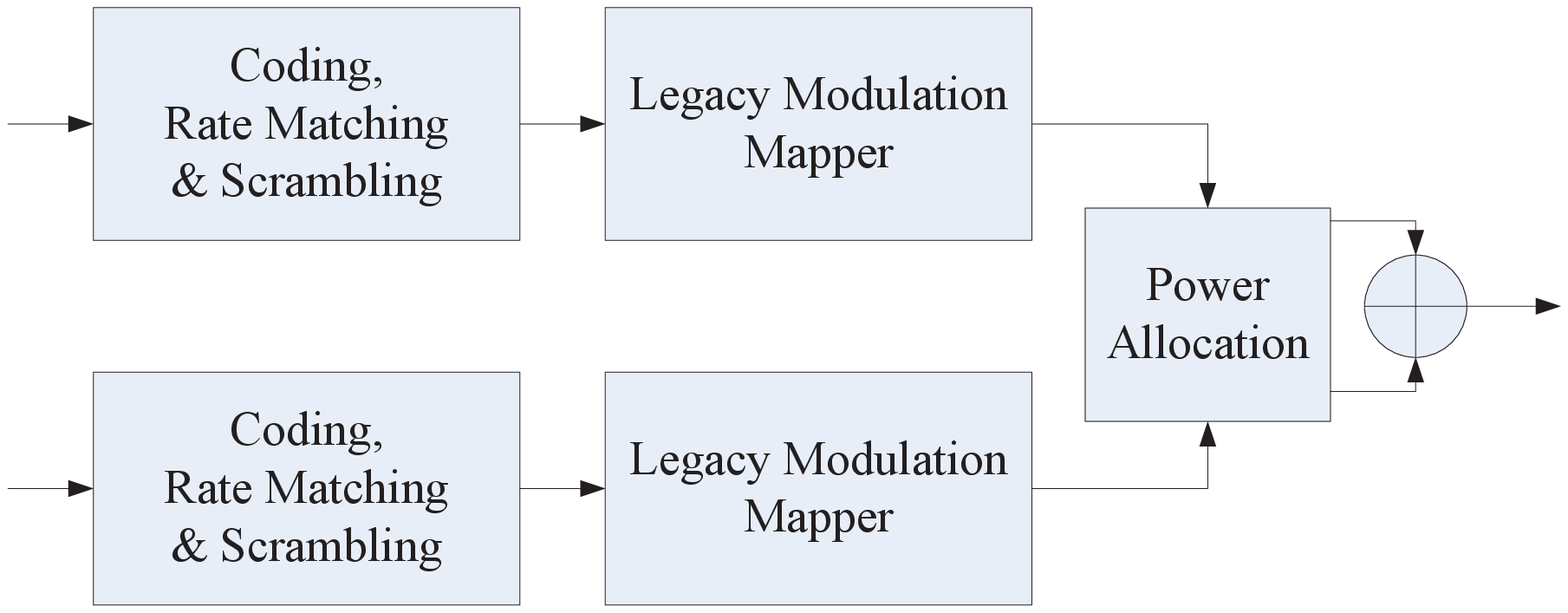}}
  \subfigure[MUST Category 2]{
  \includegraphics[width=0.42\textwidth]{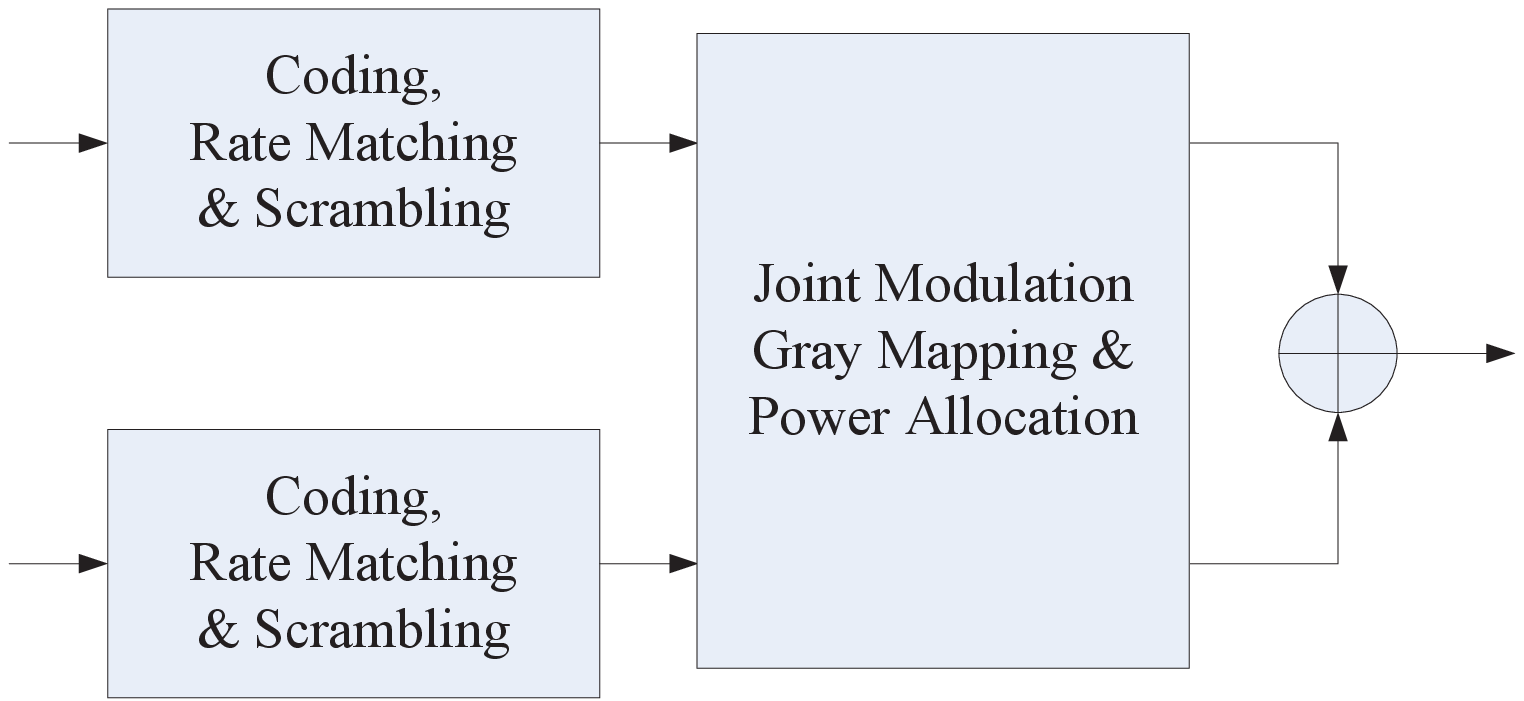}}
  \subfigure[MUST Category 3]{
  \includegraphics[width=0.5\textwidth]{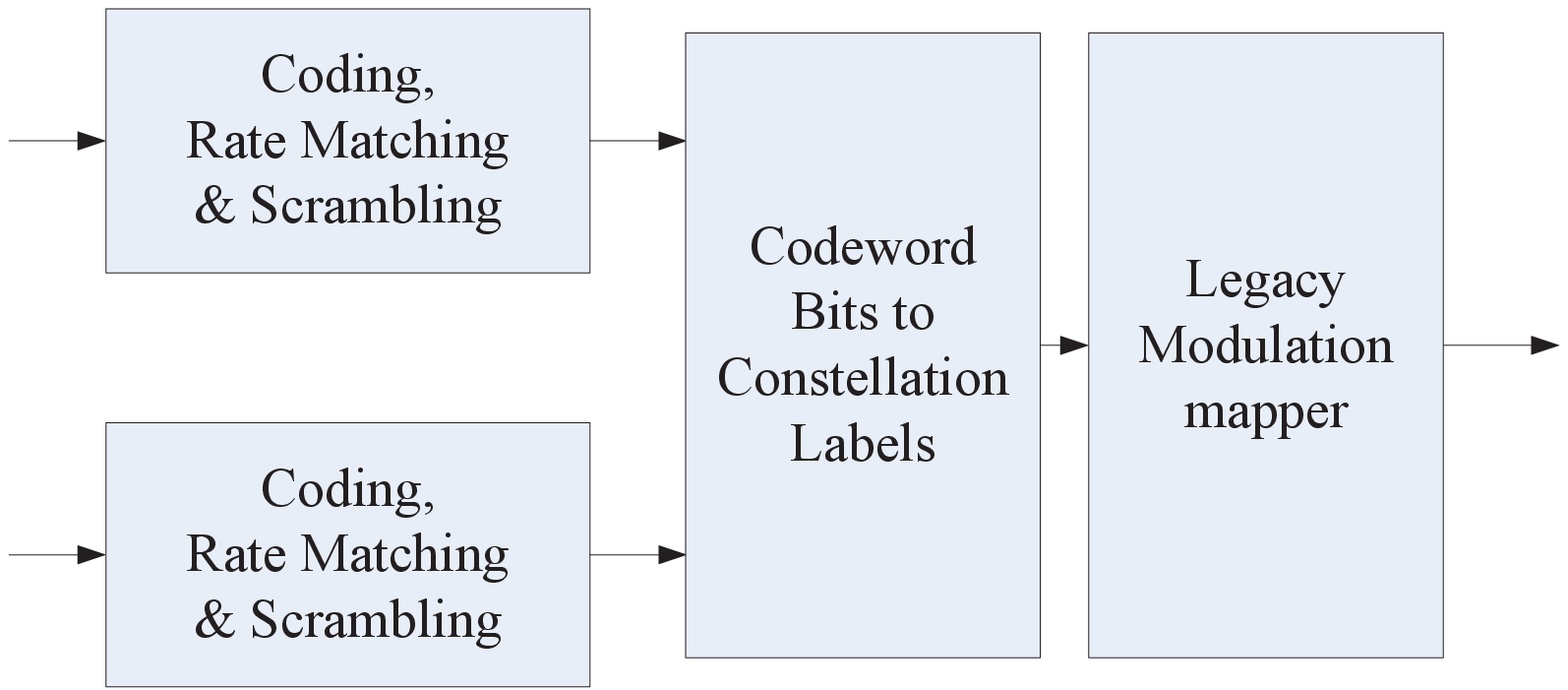}}
   \caption{Examples of transmitter processing of candidate MUST schemes}
     \label{fig1}
\end{figure}

There have been a number of standardization activities related to the implementation of NOMA in next-generation mobile networks.  In particular, the   standardization organization 3GPP initiated a study item on Downlink MUST for LTE in Release 13, focusing on multiuser non-orthogonal transmission schemes, advanced receiver designs and related signaling schemes \cite{MUST}. Various non-orthogonal transmission schemes have been  proposed and studied in the MUST study item. Based on their characteristics, they can be generally divided  into three categories \cite{scheme1-1}, and examples of transmitter processing for these three categories are shown in Fig. \ref{fig1}.
\begin{enumerate}
\item Category 1: Superposition transmission with an adaptive power ratio on each component constellation and non-Gray-mapped composite constellation.
\item Category 2: Superposition transmission with an adaptive power ratio on component constellations and Gray-mapped composite constellation.
\item Category 3: Superposition transmission with a label-bit assignment on composite constellation and Gray-mapped composite constellation.
\end{enumerate}

To characterize the gains of the non-orthogonal transmission schemes studied in MUST quantitatively, the initial link level and system level evaluation has been  provided by various companies. It is envisioned that almost $20$\% cell-average and cell-edge throughput gains can be obtained \cite{scheme1-1}. Other topics supporting the non-orthogonal transmission, e.g.,  channel state information (CSI) reporting schemes, retransmission schemes, HARQ process design and the signaling schemes associated with the advanced receiver, are still under active discussion.

In addition to MUST, there are other forms of non-orthogonal multiple access schemes, e.g., sparse code multiple access (SCMA), pattern division multiple access (PDMA), multiuser shared multiple access (MUSA), which have also been actively studied as promising MA technologies  for 5G \cite{Whitepaper1, Whitepaper2}:
\begin{enumerate}
\item SCMA is proposed as a multi-dimensional constellation codebook design based on the non-orthogonal spreading technique, which can be overloaded to enable massive connectivity and support grant-free access. SCMA directly maps the bit streams to different sparse codewords, and different codewords for all users are multiplexed over shared orthogonal resources, e.g., OFDM subcarriers. At the receiver, a low-complexity message passing algorithm is utilized to detect the users' data.

\item The uplink MUSA scheme is based on the enhanced multi-carrier code division multiple access (MC-CDMA) scheme. Equipped with advanced low correlation spreading sequences (e.g., I/Q data randomly taking $\{-1,0,1\}$ values at the transmitter),  linear processing and SIC techniques at the receiver, MUSA can achieve remarkable gains in system performance, especially when the user overloading factor is high, for example, larger than $300$\%. 

\item PDMA employs  multiple-domain non-orthogonal patterns, which are realized  by maximizing the diversity and minimizing the overlaps among multiple users. The multiplexing can be realized in the code, power, and spatial domains or their combinations, which enables high flexibility for coding and decoding processing. PDMA can promote 1-2 times increase of the system spectral efficiency, decrease data transmission delay and enhance quality of experience (QoE) of user access.
\end{enumerate}

It is worth pointing out that these aforementioned MA candidates are closely related to the fundamental principle of NOMA which is to serve multiple users at the same channel use. Take SCMA  as an exmaple. The term of sparsity refers to the fact that each user can  occupy only a small number of orthogonal channel uses, such as subcarriers, but   there is always more than one user occupying each of the subcarriers. Therefore, at each subcarrier, SCMA can be viewed as NOMA, since multiple users are sharing the same bandwidth resource. Or in other words, SCMA can be built by combining NOMA with advanced strategies for subcarrier allocation, coding and modulation.

\section{Research Challenges}
\subsubsection{User Pairing/Clustering}
 While most of the examples provided in this article consider  two-user downlink scenarios, it is important to point out that NOMA can be applied to   general uplink and downlink scenarios with more than two users.  However, the use of superposition coding and SIC can cause extra system complexity, which motivates the use of user pairing/clustering,  an effective approach to reduce the system complexity since fewer users are coordinated for the implementation of NOMA.   However, in cluster-based NOMA systems, it is very challenging to determine  how best to dynamically allocate users to a fixed/dynamic number of clusters with different sizes. It is important to point out that the resulting combinatorial optimization problem is in general NP-hard and performing an exhaustive search for an optimal solution is computationally prohibitive.   Therefore,  it is important to propose new low-complexity algorithms to realize optimal user clustering.  Note that the performance of the cluster-based NOMA system can be further improved by opportunistically performing power allocation among different users in each cluster.

\subsubsection{Hybrid Multiple Access}
It has been envisioned that future cellular networks will be designed by using more than one MA techniques, and this trend  has also been evidenced by the recent application of NOMA to 3GPP-LTE (MUST). Particularly,   MUST is  a hybrid MA scheme between OFDMA and NOMA, where NOMA is to be used when users have very different CSI, e.g., one user close to the BS and the other at the cell edge.  Therefore it is important to study how to combine NOMA with other types of MA schemes, including not only the conventional OMA schemes  but also those newly developed 5G MA technics.  Advanced game theoretic approaches can be applied to optimize the use of bandwidth resources in the power, frequency, time and code domains.

\subsubsection{MIMO-NOMA}
In NOMA with beamforming, there are still various issues and
challenges.
For example, optimal joint user allocation
and beamforming schemes have not been considered as their computational
complexity would be prohibitively high.
Joint transmit and receive beamforming is also an important
topic that has not been well investigated yet.
The main difficulty for NOMA with spatial multiplexing  is the complexity of the receivers of the users.
A strong user needs to jointly detect multiple signals twice, which
might be computationally demanding.
The extension of
NOMA with spatial multiplexing to more than two
users with multiple carriers
also requires user clustering and resource allocation
in a multi-dimensional space (i.e., frequency, time,
spatial, and power domains), which is an analytical and computational challenge.

\subsubsection{Imperfect CSI}
Most existing work on NOMA has relied on the  perfect CSI  assumption  which is difficult to realize, since sending more pilot signals  to improve the accuracy of channel estimation reduces the  spectral efficiency.  Therefore, it is important to study the impact of   imperfect CSI on the   reception reliability in NOMA systems. Another example of the strong CSI assumptions is that  many NOMA protocols require the CSI at the transmitter, which can cause significant system overhead.  The use of only a few bits of feedback is a promising solution in NOMA systems, since obtaining the ordering  of users' channel conditions is sufficient for the implementation of NOMA in many applications.

\subsubsection{Cross-layer Optimization}
Cross-layer optimization is important to maximize the performance of NOMA in practice and  meet the diversified  demands of 5G, e.g, spectral efficiency, energy efficiency, massive connectivity, and low latency. For example, practical designs of coding and modulation are important to realize the performance gain of NOMA at the physical layer, and it is crucial to study how to feed these gains from the physical layer to the design of upper layer protocols. This cross-layer optimization is particularly important to NOMA which, unlike conventional MA schemes,  takes the user fairness into consideration, which means that the issues related to user scheduling and pairing, power allocation,  pilot and retransmission schemers design need to be jointly optimized.

\section{Conclusions}
In this article, the concept of NOMA has been first illustrated by using a simple scenario with two single-antenna users. Then various forms of  MIMO-NOMA transmission protocols, the design of cooperative NOMA, and the interplay between two 5G technologies, NOMA and cognitive radio, were discussed. The recent industrial efforts for the standardization of NOMA in LTE and 5G networks have been conclusively identified, followed by a detailed discussion of research challenges and potential solutions.

\begin{IEEEbiography}{Zhiguo Ding  }    (z.ding@lancaster.ac.uk) received his Ph.D. degree  from Imperial College London in 2005, and is currently a chair professor in Lancaster University, UK.  His research interests include 5G communications, MIMO and relaying networks, and energy harvesting.   He serves as an Editor for several journals including {\it IEEE Transactions on  Communications}, {\it IEEE Communication Letters}, {\it IEEE Wireless Communication Letters} and {\it Wireless Communications and Mobile Computing}.
 \end{IEEEbiography}

\begin{biography}{
Yuanwei Liu}  (yuanwei.liu@qmul.ac.uk) is currently working toward the Ph.D. degree in Electronic Engineering at Queen Mary
University of London. Before that, he received his M.S. degree and B.S. degree from the Beijing University of Posts and Telecommunications in 2014 and 2011, respectively.

His research interests include non-orthogonal multiple access, wireless energy harvesting, Massive MIMO, Hetnets, D2D communication, cognitive radio, and physical layer security. He received the Exemplary Reviewer Certificate of the IEEE Wireless Communication Letter in 2015. He has served as TPC member for IEEE conferences such as IEEE GLOBECOM.
\end{biography}

\begin{biography}
{Jinho Choi}  (jchoi0114@gist.ac.kr)   received his Ph.D. degree from KAIST, Korea, in 1994, and is with Gwangju Institute of Science and Technology (GIST), Korea, as a professor. His research interests include statistical signal processing and wireless communications. He authored two books (``Adaptive and Iterative Signal Processing in Communications" and ``Optimal Combining and Detection") published by Cambridge University Press and currently serves as an Editor for {\it IEEE Transactions on Communications}.
\end{biography}

\begin{biography}
{Qi Sun} (sunqiyjy@chinamobile.com) received her Ph.D. degree in information and communication engineering from Beijing University of Posts and Telecommunications in 2014. After graduation, she joined the Green Communication Research Center of the China Mobile Research Institute. Her research interest focuses on 5G communications, including new waveforms, non-orthogonal multiple access, massive MIMO and full duplex.
\end{biography}

\begin{biography}
{Maged Elkashlan}(maged.elkashlan@qmul.ac.uk) received the Ph.D. degree in Electrical Engineering from the University of British Columbia in 2006. From 2007 to 2011, he was with the Wireless and Networking Technologies Laboratory at Commonwealth Scientific and Industrial Research Organization (CSIRO), Australia. In 2011, he joined the School of Electronic Engineering and Computer Science at Queen Mary University of London. Dr. Elkashlan serves as Editor of \textsc{IEEE Transactions on Wireless Communications}, \textsc{IEEE Transactions on Vehicular Technology}, and \textsc{IEEE Communications Letters}.
\end{biography}

\begin{biography}
{Chih-Lin I} (icl@chinamobile.com) is CMCC Chief Scientist of Wireless Technologies, launched 5G R$\&$D in 2011, and leads C-RAN, Green and Soft initiatives. Chih-Lin received IEEE Trans. COM Stephen Rice Best Paper Award, and IEEE ComSoc Industrial Innovation Award. She was on IEEE ComSoc Board, GreenTouch Executive Board, WWRF Steering Board, M$\&$C Board Chair, and WCNC SC Founding Chair. She is on IEEE ComSoc SPC and EDB, ETSI/NFV NOC, and Singapore NRF SAB.
\end{biography}

 \begin{IEEEbiography}{H. Vincent Poor} (poor@princeton.edu) is with Princeton University, where his interests are in wireless networking and related fields. He is a member of the National Academy of Engineering and the National Academy of Sciences, and a foreign member of the Royal Society. He received the IEEE ComSoc Marconi and Armstrong Awards in 2007 and 2009, respectively, and more recently the 2016 John Fritz Medal and honorary doctorates from several universities.
 \end{IEEEbiography}

 \end{document}